\documentclass[prl,twocolumn,showpacs]{revtex4}
\usepackage[xdvi]{graphicx}
\usepackage{latexsym}
\usepackage{amsbsy}
\usepackage{amsmath}
\usepackage{times}
\usepackage{verbatim}
\renewcommand{\vec}[1]{\mbox{\boldmath$\mathrm{#1}$}}

\begin{document}

\date{\today}
\title{Local control  of  ultrafast dynamics of  magnetic nanoparticles}

\author{ A. Sukhov$^{1, 2}$ and J. Berakdar$^2$}
\address{$^1$Max-Planck-Institut f\"ur Mikrostrukturphysik,
Weinberg 2, D-06120 Halle/Saale, Germany\\
$^2$Institut f\"ur Physik, Martin-Luther-Universit\"at
 Halle-Wittenberg, Heinrich-Damerow-Str. 4, 06120 Halle, Germany}
\begin{abstract}
Using the local control theory we derive analytical expressions for magnetic field pulses
that steer the magnetization
of a monodomain magnetic nanoparticle  to a predefined state.
 Finite-temperature full numerical
simulations confirm the analytical results and show that a
magnetization switching or freezing is achievable within
few precessional periods and that the scheme is exploitable for fast thermal switching.
\end{abstract}
\maketitle
\begin{center}
   \begin{figure}[h]
    \centering  \includegraphics[width=.34\textwidth]{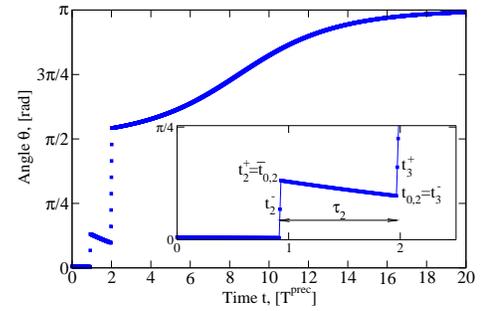}
	     \caption{\label{fig_1} (Color online) Evolution of $\theta(t)$ according to the proposed control scheme and	  for     $\phi(t=0)=\pi/180=\theta(t=0),\phi_0=\arctan (b_y/b_x)=2 \pi/3,\,\alpha=0.05,\, f =1,\, b_0=0.2$.
	     Inset shows the short-time behaviour (pulses are off for $\theta>\pi/2$).}
    \end{figure}
\end{center}
\begin{center}
   \begin{figure}[h]
    \centering  \includegraphics[width=.34\textwidth]{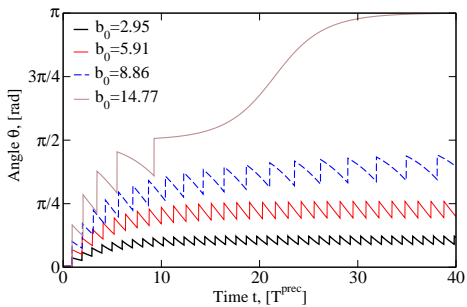}
	     \caption{\label{fig_3} (Color online) Tilt angle $\theta(t)$
	     within the present the local control scheme for different
fields strengths $b_0$. Other parameters: $\alpha=0.05$, $T_0=0 K$. (Pulses are off when $\theta>\pi/2$).}
    \end{figure}
\end{center}
\begin{center}
   \begin{figure}[h]
\vspace{3ex}
    \centering  \includegraphics[width=.34\textwidth]{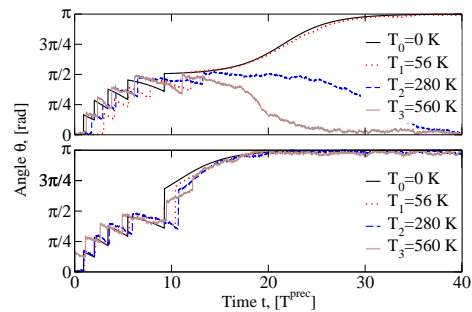}
	     \caption{\label{fig_4} (Color online) Temperature-dependent controlled evolution of the angle
	     $\theta(t)$ ($\alpha=0.05$, $b_0=14.77$).  The pulses are applied
	     if $\theta<\pi/2$ only (top panel) or  throughout (below).}
    \end{figure}
\end{center}
\begin{center}
   \begin{figure}[h]
    \centering  \includegraphics[width=.36\textwidth]{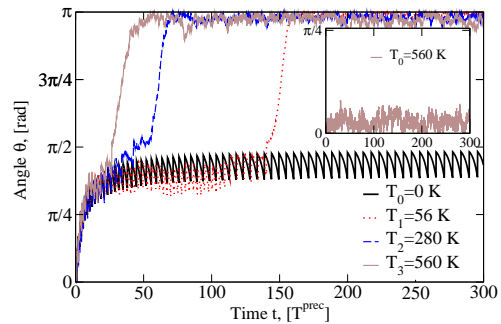}
	     \caption{\label{fig_5} (Color online) Thermal-assisted controlled switching in the presence
	     of short pulses with an amplitude $b_0=8.86$. Inset shows
	     switching is not possible for $b_0=0$.}
    \end{figure}
\end{center}

\emph{Introduction.-}
A fast magnetization reversal of magnetic nanoparticles is of a key
importance for the realization of high-rate  magnetic recording
 \cite{HiOu01,HiOu0306}. Several techniques are currently
 envisaged for the magnetization switching
   such as  the laser-induced    spin dynamics \cite{Vomi05} based on   the
inverse Faraday effect \cite{Kimel06, Perr06},  the reversal
triggered by external static or alternating magnetic fields
\cite{StonWohl48,chantrell,Thir03,Gerrits,woltersdorf,noz,Garzon} or by  a spin-torque
 acting on the magnetization due to a passing spin polarized electric current \cite{Slon96, Berg96}.
 Transverse magnetic field pulses are also efficient
 for a swift reversal  \cite{He96,Back98,Back99,Bauer00,Xi08,SuWa06},  and if finely tuned in duration \cite{Schuh03,HiOu01}
 can even lead to a quasi-ballistic switching.
A further fundamental issue, addressed here is how to steer the magnetic
dynamics to a desirable state by external fields. Generally,  a number of control schemes have been established
 mainly in quantum chemistry \cite{18,19,20,21}. Particularly interesting is the
local control theory (LCT) \cite{20,21} in which
 the control fields are constructed from the response of the system offering thus
 a physical interpretation of the control mechanism.
We adopt the  idea of  LCT
 to steer the magnetization dynamics of
nanoparticle by transverse magnetic pulses.  We obtain  transparent analytical expressions
 for the  control  pulses that allow
a fast switching or a quasi "freezing" at a predefined magnetization state.
 For the scheme to be applicable, the field durations have to be shorter
than the field-free precessional period but no special pulse-duration tuning is required;  the
field strengths are to be determined according to the analytical expressions provided here.
In our control strategy  the magnetization dynamics proceeds via sudden impulsive kicks guiding the magnetization towards a predefined direction; the pulses are intervened by field-free magnetization precessions and relaxation. A similar mechanism has recently been realized experimentally \cite{Garzon} using spin-polarized
picosecond current pulses resulting in a spin-transfer-torque-driven stroboscopic dynamics.
 The robustness of the predictions we demonstrate with finite-temperature full numerical calculations
  and for  different types of anisotropy fields. We
confirm the analytical results and uncover the potential of this scheme for fast thermal
switching that can be the basis for fast thermal sensors. \\
\indent \emph{Theory.-}
We consider a nanoparticle  with a size such that it
displays a long-range magnetic order and is in a single domain remanent state.
 Examples are Fe$_{50}$Pt$_{50}$ \cite{HiOu01, Anto06}
or  Fe$_{70}$Pt$_{30}$ \cite{HiOu01, Anto05} nanoparticles
which possess respectively a uniaxial or a cubic anisotropy.
Following the  Landau-Lifshitz-Gilbert (LLG) approach we model
the   dynamics of the magnetization direction  by the classical evolution
of a unit vector $\mathbf{S}$. The particle's magnetic moment at saturation $\mu_S$
is assumed time-invariant. The system energy derives from
$
\displaystyle \mathcal{H}\,=\mathcal{H}_A + \mathcal{H}_F,
$
where $\mathcal{H}_A$ and $\mathcal{H}_F=-\vec{S}\cdot\vec{b_0}(t)$ stand respectively  for the anisotropy
 and the Zeeman energy of $\vec{S}$ in the external field $\vec{b_0}(t)$.
For a particular type of anisotropy described by  $f_A(\mathbf{S})$ we write
$\mathcal{H}_A=-D f_A(\mathbf{S})$ with $D$ being the anisotropy
constant. $\mathbf{S} (t)$ develops according to LLG  equation \cite{LaLi35} as
$\frac{\partial {\mathbf{S}}}{\partial t}
\,=\,-\,\frac{\gamma}{(1+\alpha^2)}{\mathbf{S}}
 \times \Big[{\mathbf{B}_{e}}(t)\,+\,\alpha \big ({\mathbf{S}}
  \times {\mathbf{B}_{e}}(t) \big) \Big],
$
where   ${\mathbf B}_{e}(t)
 = - 1/(\mu_S)\partial \mathcal {H} /\partial {\mathbf S}$ is the effective field,
$\gamma$ is the gyromagnetic ratio and
$\alpha$ is the  Gilbert damping parameter.
 In  spherical coordinates where the $z$ axis along is the easy axis we  specify  $\mathbf S$ by the azimuthal
 ($\phi$) and polar
($\theta$) angles
 and cast  the LLG equation as \cite{HiOu01,Vons66}
\begin{equation}
     \begin{array}{l}
       \displaystyle{(1+\alpha^2)\frac{d\phi}{dt}=\frac{1}{\sin \theta} \, \frac{\partial \mathcal{H}}{\partial \theta}\,-\,\frac{\alpha}{\sin^2 \theta}\, \frac{\partial \mathcal{H}}{\partial \phi},} \\
       \displaystyle{(1+\alpha^2)\frac{d\theta}{dt}=-\frac{1}{\sin \theta} \, \frac{\partial \mathcal{H}}{\partial \phi}\,-\,\alpha \, \frac{\partial \mathcal{H}}{\partial \theta}.}
     \end{array}
     \label{eq_1}
   \end{equation}
  Hereafter the time is
measured  in  units of the field-free precessional period $T^{\mathrm{prec}}$ and
the energy ${\cal H}$
 in units of  $\mu_S B_A$ where $B_A=2D/\mu_S$  is the maximum
uniaxial anisotropy field. E.g., for
   Fe$_{50}$Pt$_{50}$ we have $T^{\mathrm{prec}}= 5$ ps, the maximum anisotropy field is $\sim$ 7 T and the magnetic moment per
nanoparticle is around $22000\mu_{\mathrm{B}}$ \cite{Anto06}.
The field-free solution of (\ref{eq_1}) is  known; e.g. for a uniaxial anisotropy and
starting from the angles $\phi_f(t=\bar t_0)$ and $\theta_f(t=\bar t_0)$ one finds (e.g., \cite{Garcia})
\begin{equation}
     \begin{array}{l}
       \displaystyle{\phi_f(t)=\phi_f
       (\bar t_0)\pm\frac{t-\bar t_0 }{1+\alpha^2}\pm}\\
       \displaystyle{\frac{1}{\alpha}\ln\left|\frac{\cos\theta_f(\bar t_0)(1+\sqrt{1+\tan^2\theta_f(\bar t_0)
       \cdot\mathrm{e}^{-\frac{2\alpha
       (t-\bar t_0)}{1+\alpha^2}}})}{1+\cos\theta_f(\bar t_0)}\right|,} \\
       \displaystyle{\tan\theta_f(t)=\tan\theta_f (\bar t_0) \cdot \mathrm{e}^{-\frac{\alpha}{1+\alpha^2}(t-\bar t_0)}.}
     \end{array}
     \label{eq_6}
   \end{equation}
``+'' (``-'') refers to $0<\theta<\pi/2\,$  ($\,\pi/2<\theta<\pi$).\\
To control the dynamics we apply along the $x$ and $y$ axis
two  magnetic field pulses $\textbf{b}_{\mathrm{x}}$ and $\textbf{b}_{\mathrm{y}}$
of durations $2\varepsilon$ and shapes $f(t)$  centered at  some moment $t=t_0$.
Their relative strengths is given by the mock angle $\phi_0$, with $\tan \phi_0=|b_{\mathrm{y}}|/|b_{\mathrm{x}}|$;  the total
fields strength is  $|f|b_0/(2\varepsilon)$. Hence  $\vec{b}_{0}(t)=\mathbf b_x+\mathbf b_y$ is

\begin{equation}\label{eq:field}
       \vec{b}_{0}(t)=\left\{\begin{array}{ll}
                              \frac{f(t)b_0}{2\varepsilon} (\cos \phi_0 \vec{e}_{\mathrm{x}}+\sin \phi_0 \vec{e}_{\mathrm{y}}), & t_0-\varepsilon<t<t_0+\varepsilon\\
                               0, &\mathrm{elsewhere.}
                                 \end{array} \right.
       \end{equation}
Switching to a new time variable
$\tau(t)=\frac{t-(t_0+\varepsilon)+2\varepsilon}{2\varepsilon}$
we derive for the equation of motion
\begin{equation}
     \begin{array}{l}
       \displaystyle{\frac{1}{2\varepsilon}\frac{d\phi}{d\tau}=\hspace{1ex}p\left[\frac{1}{\sin\theta}\frac{\partial
       \mathcal{H}_A}{\partial
       \theta}-\frac{\alpha}{\sin^2\theta}\frac{\partial
       \mathcal{H}_A}{\partial \phi} \right]}\\
          \displaystyle{\hspace{8ex}-\frac{pb_0 f(t(\tau))}{2\varepsilon}\left[\frac{\cos\theta}{\sin\theta}\cos\delta\phi+\alpha\frac{\sin\delta\phi}{\sin\theta}\right],} \\
       \displaystyle{\frac{1}{2\varepsilon}\frac{d\theta}{d\tau}=\hspace{1ex}p\left[-\frac{1}{\sin
       \theta} \frac{\partial \mathcal{H}_A}{\partial \phi}\,-\,\alpha
       \frac{\partial \mathcal{H}_A}{\partial \theta}\right]}\\
   \displaystyle{\hspace{8ex}+\frac{pb_0 f(t(\tau))}{2\varepsilon}\left[-\sin\delta\phi+\alpha\cos\theta\cos\delta\phi\right],}
     \end{array}
     \label{eq_5}
   \end{equation}
where  $\delta\phi=\phi-\phi_0$ and $p=1/(1+\alpha^2)$.
If the magnetic pulses are shorter than the precessional period then from eq. (\ref{eq_5})
we infer for the angles stroboscopic evolution
   from before $\left( \phi(t^-), \theta(t^-)\right)$ to after $\left(\phi(t^+), \theta(t^+)\right)$
     the pulses
the relation  (we introduced$\, t^-:=t_0-\varepsilon,\, t^+:=t_0+\varepsilon$)
\begin{equation}
     \begin{array}{l}
       \displaystyle{\frac{d\phi}{d\tau}=-\frac{1}{\sin\theta}\frac{b_0 f(t_0)}{1+\alpha^2}\left[\cos\theta\cos\delta
       \phi+\alpha\sin\delta \phi\right],} \\
       \displaystyle{\frac{d\theta}{d\tau}=\frac{b_0f(t_0)}{1+\alpha^2}\left[-\sin\delta
       \phi+\alpha\cos\theta\cos\delta \phi\right],}
     \end{array}
     \label{eq_7}
   \end{equation}
  which is valid up to terms of the order $(\epsilon/T^{\rm prec})^2$.
   After the pulse, i.e. for  $t> t^+$ the dynamics is governed by eq.(\ref{eq_6}) with the
   initial conditions  $\phi_f=\phi(t^+)$, $\theta_f=\theta(t^+)$. This procedure is   repeated
   by applying further pulses accordingly.\\
\indent \emph{Controlled switching.- }  As we are interested in switching we require in the spirit of local control theory that
\begin{equation}
\theta(t^+)>   \theta(t^-)\quad \forall\, t^+, t^-.
\end{equation}
As inferred from eq.(\ref{eq_7}), this condition is fulfilled if $\delta \phi=\phi-\phi_0=3\pi/2$.
If a  sequence   of  the pulses (\ref{eq:field})  each centered
at the times $t_{0,i}$ is applied  then  $\mathbf {S}(t)$  evolves as
\begin{equation}
     \begin{array}{l}
       \displaystyle{\phi{(t^+_i)}=\phi{(t^-_i)}+\alpha \ln\left|\frac{\tan\left(\frac{\theta{(t^-_i)}}{2}+\frac{1}{2}
       \frac{b_0f(t_{0,i})}{1+\alpha^2}\right)}{\tan\left(\frac{\theta{(t^-_i)}}{2}\right)}\right|,} \\
       \displaystyle{\theta{(t^+_i)}=\theta{(t^-_i)}+\frac{b_0f(t_{0,i})}{1+\alpha^2},}
     \end{array}
     \label{eq_9}
\end{equation}
where $t^\pm_i=t_{0,i}\pm\varepsilon$.\\
The realization of this LCT  scheme is then as follows: Starting from a known
(e.g. equilibrium) state $\phi=\phi(0)$;
$\theta=\theta(0)$
we apply at $t=t_{0,1}$ the first fields $b_x$ and $b_y$ (\ref{eq:field}) with strengths such that
$\phi_0=\phi(0)-3\pi/2$ (cf. Fig.(\ref{fig_1})). Eq.~(\ref{eq_9}) delivers  the tilt angles $\theta(t^+_1)$ and $\phi(t^+_1)$.
During a time lag (dark time) $\tau_1$ the propagation proceeds according to eq.(\ref{eq_6}) with the initial values
$\phi_f(\bar t_0)=\phi(t^+_1)$ and $\theta_f(\bar t_0)=\theta(t^+_1)$. At $t=t_{0,2}$ we apply a second pulse
with $b_x$ and $b_y$ such that $\phi_0=\phi_f(t^+_1+\tau_1)-3\pi/2$. From eq. (\ref{eq_9}) we deduce
that after the second pulse $\theta(t_2^+)=\theta_f(t^+_1+\tau_1)+\frac{b_0f(t_{0,2})}{1+\alpha^2}$. This procedure is repeated
until we achieve the state with $\theta=\pi/2$. As clear from (\ref{eq_9}) the tilt angle is always increased upon the pulse
with an amount that goes linearly with the fields strength  $b_0$.
On the other hand the variation of  $\phi$ with $b_0$ is only logarithmic, in fact
if the time delay between the pulses is only a fraction of the precessional period, $\phi$ is hardly changed.\\
\indent \emph{Freezing.-} The present LCT scheme allows also for the stabilization of
the magnetization around  a desirable target non-equilibrium angle $\theta_t$: At first, starting from a given state we apply the control scheme and  achieve $\theta_t$ at some time $t_t$. During a field-free period  $ \tau$  the angle  $\theta_t$  develops to
$\theta_f(t_t+\tau)$. To compensate for this change
we apply  a pulse (centered at $t_{0,t}$) according to our scheme this shifts the  angle to $\theta^+=\theta_f(t_t+\tau)+\frac{b_0f(t_{0,t})}{1+\alpha^2}$.
To stabilize the magnetization  we choose $b_0$ such that $\theta^+=\theta_t$. The procedure is then
repeated during the stabilization time.  To minimize the adjustment
of  $b_0$  between
consequent pulses the  repetition rate should be large.
\begin{center}
   \begin{figure}[h]
    \centering  \includegraphics[width=.36\textwidth]{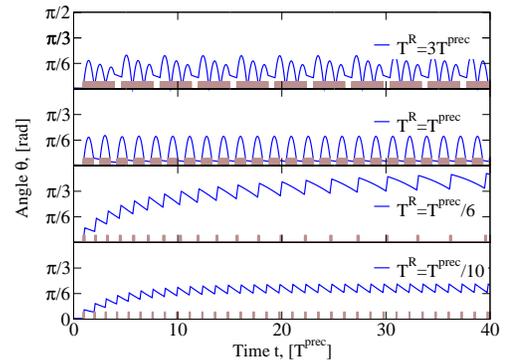}
	     \caption{\label{fig_2} (Color online)  $\theta(t)$
	     for different pulse durations
(solid rectangles). $T^{\rm prec}$ is  the
	     precessional period and $b_0=0.3$, $\alpha=0.05$.}
    \end{figure}
\end{center}
\indent \emph{Numerical results and illustrations.-}
Fig. \ref{fig_1} shows the magnetization reversal according to our  zero temperatures ($T=0$)
 analytical scheme
and in the damping regime appropriate for magnetic
nanoparticles.  Fig. \ref{fig_1} confirms our analysis and the physical picture
drawn above.  However,
the following issues need to be clarified
 for this procedure to be of practical interest.
  1.) Do we need a precise tuning of the pulses durations,
  2.) will  thermal fluctuations invalidate our findings,  and 3.)
how effective is this scheme when applied to  other type of anisotropy fields.
To address these points we implemented a finite-temperature
full numerical realization \cite{SuBer08} of the present  control scheme (cf.~\cite{HiOu01,HiOu0306} and references therein for an overview on numerical micromagnetic methods), i.e.
the analytical expressions deliver the appropriate
 input parameters  for the numerics. The damping parameter  is
chosen according to experimental findings \cite{HiOu01}.
For the simulation presented here we use square-shaped pulses, i.e. $f(t)=1$
for $t_0-\varepsilon<t<t_0+\varepsilon$.
Basically the same conclusions are valid for other pulse shapes, e.g. Gaussian
pulses \cite{alex_unpub}.
Fig. \ref{fig_2} demonstrates the evolution sensitivity of  the angle $\theta$ when pulses
  with different durations are applied. It also shows the range
   of validity of our scheme. As inferred from
   Fig. \ref{fig_2}  a fine tuning of the pulse duration is
   not mandatory as long as it is smaller that  $T^{\mathrm{prec}}$.
   The strength $b_0$  determines the value  of the tilt angle (as follows from eq.(\ref{eq_9})).
  The insensitivity to the pulse duration is favorable for  practical
  applications, however the generation of magnetic pulses shorter than $T^{\mathrm{prec}}$ might be
    a challenge; the light-induced  generation of sub-picosecond shaped
    magnetic pulses \cite{andrey} may circumvent this problem.
As for the role of the magnetization dynamics during the pulses  our full numerical
 simulations (cf. Fig. \ref{fig_3}) confirm qualitatively the
  analytical predictions.
 According to eq.(\ref{eq_9}) a minimal fields strength $b_0$ is required for switching, for  $b_0$ determines
$\theta(t^+)$.  To realize the stabilization scheme outlined above
   one tunes $b_0$ to steer the magnetization to a non-equilibrium   $\theta_t$  (cf. Fig. \ref{fig_3}) and
   keep it there (as long as $\mathbf b_0$ is on).
Figure \ref{fig_4} proves the robustness of the scheme to  thermal
fluctuations.
Here we highlight  a special feature of the temperature-dependent magnetization dynamics:
To achieve switching, the pulses have to be applied even if
$\theta_t>\pi/2$, since due to thermal excitations the
magnetization may swing back to the original state. This effect is
avoided by applying the pulses even if $\theta>\pi/2$ (Fig. \ref{fig_4}, lower
panel). Generally, we observe that thermal fluctuations have little influence on the
effect of the pulses (i.e., on the dynamics during and right after the pulses),
in contrast to continuous fields \cite{SuBer08}. On the other hand
the field-free processional motion between the pulses is generally  modified at $T>0$.\\
The possibility of field-assisted stabilization (freezing) can be exploited for
fast field-assisted thermal switching: Starting at $T\approx 0$ we utilize our scheme to
 drive the magnetization  to a state   $ \theta_t\lesssim \pi/2$ (as shown in Fig. \ref{fig_5}) and then freeze it there.
  At low temperatures  switching does not occur irrespective of the
  waiting time (inset of Fig. \ref{fig_5}). When the
   temperature  increases however, the  thermal fluctuations increase but can not lead
   to a reversal in absence of the field, as demonstrated  by the inset of Fig. \ref{fig_5}.
   The presence of the fields assists
    a fast magnetization reversal, a behaviour that can not be realized with
    static fields, since a magnetization freezing is necessary.
    In practice,  the reversal process may   be functionalized a fast thermal sensor to monitor swiftly
    a temperature increase.
The question of to what extent the present scheme is applicable to
 another anisotropy type we address by studying the magnetization control of
  Fe$_{\mathrm{70}}$Pt$_{\mathrm{30}}$-nanoparticles  which possesses  cubic
  anisotropy  \cite{Anto05,Chikaz}. For a cubic anisotropy the field-free
  ground-state energy landscape contains several minima \footnote{For the cubic
 anisotropy $f_{\mathrm{A}}(\vec{S})=S^2_xS^2_y+S^2_xS^2_z+S^2_yS^2_z$.}.
  By switching we mean then a magnetization transfer between these minima
  and  not necessarily a change from a parallel to an antiparallel
  state, as in the uniaxial case.  Fig. \ref{fig_6} demonstrates the applicability
  of our control proposal. Starting from a state close to one energy
  minimum the magnetization precesses and relaxes in a field-free manner to the ground state.
  When the magnetic pulse is applied according to our LCT
  the magnetization is transferred almost directly
  to the next energy minimum in the positive energy
  semi-sphere. With the freezing scheme outlined above it is even possible to stabilize the
  magnetization on top of the barrier (Fig. \ref{fig_6}).
\emph{Summary.-}
A sequence of two perpendicular magnetic pulses, each with  a duration less than the precessional period
is  capable of increasing monotonically  the magnetization tilt angle as to achieve a predefined state.
As shown analytically (for $T=0$), this is possible if the relative strengths is tuned appropriately.
Full numerical simulations accounting for finite temperatures and different types of
anisotropy fields demonstrated the usefulness of this scheme for
 magnetization ``freezing'' and switching within tens of picoseconds.
 We also illustrated how the method
 can be  exploited for picosecond field-assisted thermal-switching.
\begin{center}
   \begin{figure}[h]
 \includegraphics[width=.2\textwidth]{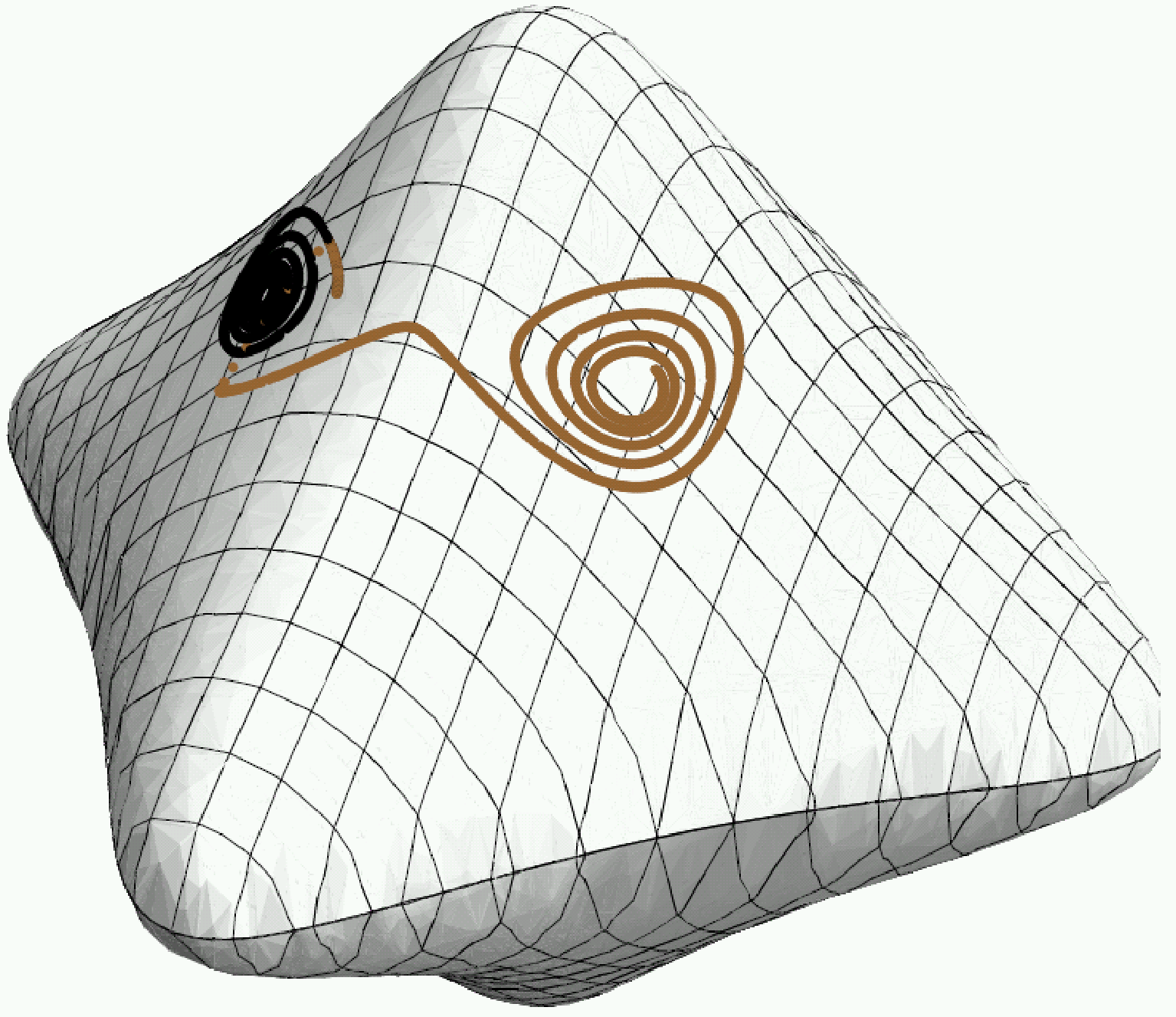}
    \includegraphics[width=.2\textwidth]{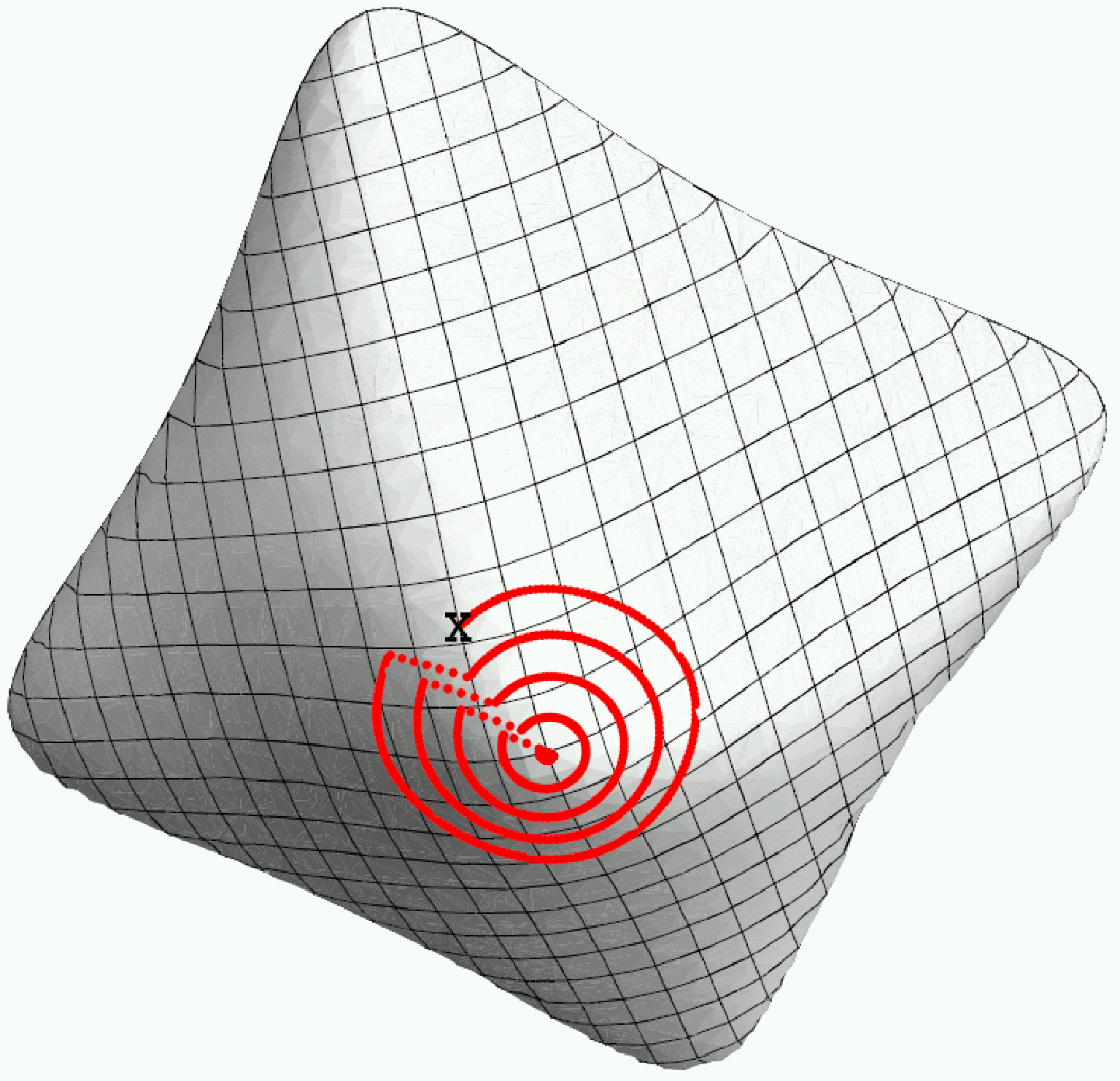}
	     \caption{\label{fig_6}
 (Color online) Polar diagram of the energy
         surface  for a cubic anisotropy with
magnetization trajectories. Left panel is a top view on the energy surface:  For $b_0=0$ (dark trajectory);
         For a $b_0=2.06$ control field  (light
trajectory).  Trajectories start at $\phi(t=0)=1.9\pi$,
         $\theta(t=0)=\pi/3.8$.
Right panel is a bottom view at the energy surface: 
        Freezing field is $b_0=0.59$ and the magnetization is initially at the position marked (X).
         In both cases
$\alpha=0.05$. 
 %
 %
 %
      }
    \end{figure}
\end{center}
%
This work is supported by the International Max Planck Research
School for Science and Technology of Nanostructures.
%

\end{document}